\documentclass[twocolumn]{aastex6}
\usepackage{natbib}
\newcommand{\lya}{Ly$\alpha$}

\newcommand{\lyb}{Ly$\beta$}

\newcommand{\lam}{$\lambda$}

\newcommand{\coXsig}{$X\,^{1}\Sigma^{+}$}
\newcommand{\coApi}{$A\,^{1}\Pi$}

\def\photofl{photons cm$^{-2}$ s$^{-1}$ \AA $^{-1}$}

\slugcomment{Accepted for publication in AJ 2016 June 17}
\shorttitle{Far-UV observations of ISON with FORTIS}
\shortauthors{McCandliss et al.}

\begin{document}


\title{Far-Ultraviolet Observations of Comet C/2012 S1 (ISON) from FORTIS}


\author{Stephan R. McCandliss\altaffilmark{1,2}, Paul D. Feldman\altaffilmark{1}, Harold Weaver\altaffilmark{3}, Brian Fleming\altaffilmark{4}, Keith Redwine\altaffilmark{1}, Mary J. Li\altaffilmark{5}, Alexander Kutyrev\altaffilmark{5}, S. Harvey Moseley\altaffilmark{5}}


\altaffiltext{1}{Department of Physics and Astronomy, The Johns Hopkins University, Baltimore, MD  21218, USA}
\altaffiltext{2}{stephan@pha.jhu.edu}
\altaffiltext{3}{Johns Hopkins University Applied Physics Laboratory, Laurel, MD  20723, USA}
\altaffiltext{4}{Center for Astrophysics and Space Astronomy, University of Colorado, Boulder, CO 80309, USA}
\altaffiltext{5}{Goddard Space Flight Center, Greenbelt, MD 20771, USA}

%\email{stephan@pha.jhu.edu}


\begin{abstract}
We have used the unique far-UV imaging capability offered by a sounding rocket borne instrument to acquire observations of C/2012 S1 (ISON) when its angular separation with respect to the sun was 26$\fdg$3, on 2013  November 20.49.  At the time of observation the comet's heliocentric distance and velocity relative to the sun were $r_{h}$~=~0.43~AU and $\dot{r}_h$~=~-62.7~km~s$^{-1}$.  \replaced{The separation between and the comet and Earth and the velocity relative to Earth were $\Delta$ = 0.84~AU and $\dot{\Delta}$~=~-4.5~km~s$^{-1}$.}{}  Images dominated by \ion{C}{1}~$\lambda$1657 and \ion{H}{1}~$\lambda$1216 were acquired over a 10$^{6}$ $\times$ 10$^{6}$ km$^2$ region.  The water production rate implied by the \lya\ observations is constrained to be $Q_{H_{2}O} \approx$ 8 $\times$ 10$^{29}$ s$^{-1}$ while the neutral carbon production rate was $Q_{C} \approx$ \replaced{3}{4} $\times$ 10$^{28}$  s$^{-1}$.  The radial profile of \ion{C}{1} was consistent with it being a dissociation product of a parent  molecule with a lifetime $\sim$ 5 $\times$ 10$^{4}$ seconds, favoring a parent other than CO.  \replaced{An upper limit on the $Q_{CO}$ production rate $<$ 5  10$^{28}$ s$^{-1}$.}{We constrain the $Q_{CO}$ production rate to 5 $^{+1.5}_{-7.5} \times$ 10$^{28}$  s$^{-1}$ with 1$\sigma$ errors derived from photon statistics.}  The upper limit on the $Q_{CO}$/$Q_{H_{2}O}$ $\lesssim$ 6\%.  

\end{abstract}





\section{Introduction}
The recent apparition of comet C/2012 S1 (ISON) presented a unique opportunity to observe a dynamically new, sungrazing, Oort cloud comet \citep{Oort:1950} prior to its reaching perihelion on 2013 November 28 2013 at a distance of only 2.7 $R_{\odot}$.   The initial gas production of an Oort cloud comet, undergoing its first passage into the inner solar system, is expected to be dominated by an excess ``frosting'' of volatile ices.  This frosting is thought to have been created over a solar lifetime ($\sim$ 4.6 Gyrs) through the bombardment of the comet's icy surface by a flux of interstellar dust, cosmic rays, ultraviolet and x-ray radiation in the nether regions between the outermost solar system and the interstellar medium (ISM) \citep{Oort:1951, Whipple:1950, Whipple:1951, Whipple:1978}.  It is further thought that if the frosting is thin enough, then the mass sublimation process, precipitated by the steady increase in the radiation environment on ingress towards the sun, will gradually reveal a surface composition that is primordial in nature.  

The ingress of ISON was closely followed by a global network of amateur and profession observers shortly after its discovery by \citet{Novski:2012}. \citet{Sekanina:2014} have prepared a comprehensive review and model of its photometric and water production behavior, starting with the pre-discovery photometry and extending beyond its total disintegration at 5.2 $R_{\odot}$ $\approx$ 3.5 hours prior to perihelion.  The observations reveal that the photometric variations and water production rates were on a seesaw cycle of coma expansion and depletion on ever shortening timescales throughout ingress.  Five cycles were identified up to 2013 November 12.9.  Disintegration was presaged by two major fragmentation events, led by a 10 fold increase in the observed water production rates between November 12.9 and 16.6  and followed by a 5 fold increase between November 19.6 and 21.6 \added{\citep{Combi:2014}}.  

\deleted{The} \added{These} water production rates were derived from observations made by the Solar Wind ANisoltropies (SWAN) \lya\ camera on the {\it SOlar and Heliospheric Observer (SOHO)} satellite, \added{which provides daily all sky coverage of \lya\ emissions with $1\degr \times 1\degr$ resolution from its vantage at Earth--Sun Lagrange point 1.  Consequently SWAN water production observations typically lag those made using small apertures, such as those from the Cryogenic Echelle Spectrometer (CSHELL) at the NASA InfraRed Telescope Facility (IRTF) acquired by \citet{DiSanti:2016}.  CSHELL observations, which were acquired through a 1\farcs0 $\times$ 3\farcs0 slit, show water production dropping by factor of $\approx$ 2.5 between November 15 and 16, coincident with a drop in the visible light curve that marked the cessation of activity from event 1 as described by \citet{Sekanina:2014}.}      

We report here on far-UV observations \added{over a $30\arcmin \times 30\arcmin$ field-of-view (FOV),} from a sounding rocket borne spectro/telescope made in the intervening period of the second event on November 20.49.  \added{They serves as a probe of water production on spatial scales intermediate to those provided by CSHELL and SWAN.}

Sounding rockets offer a unique platform for observing the far-UV emission of cometary bodies in close proximity to the sun.  The far-UV bandpass provides access to a particularly rich set of spectral diagnostics for determining the  production rates of CO, H, C, O and S.  Safety concerns for {\it HST} restrict its use to solar elongation angles of $>$ 50$\degr$, translating to  heliocentric distances of $>$ 0.766 AU \added{in the case of ISON.} \deleted{; a distance at which an Oort cloud comet is less likely to have completely lost its ISM processed layer of volatile ices.}  In contrast, sounding rocket borne instruments can use the Earth's limb to occult the sun.  The observations of ISON described here were made at an elongation of 26$\fdg$3 when the comet was at a heliocentric distance  $r_{h}$~=~0.43 AU, a heliocentric velocity of $\dot{r_h}~=~-62.7$~km~s$^{-1}$,  a separation between the Earth and comet of $\Delta$ = 0.84 AU and a relative velocity with respect to Earth of $\dot{\Delta}~=~-4.5$~km~s$^{-1}$.

\section{FORTIS Instrument Overview and Calibration}

FORTIS  \citep[Far-uv Off Rowland-circle Telescope for Imaging and Spectroscopy;][]{McCandliss:2004, McCandliss:2008, McCandliss:2010, Fleming:2011} is a 0.5 m diameter f/10 Gregorian telescope (concave primary and secondary optics) with a diffractive triaxially figured secondary, creating an on-axis imaging channel and two redundant off-axis spectral channels that share a common focal plane.   The spectral channels have a inverse linear dispersion of 20 \AA\ mm$^{-1}$.  The spectral bandpass is $\sim$ 800 -- 1800 \AA.  The imaging plate scale is 41$\farcs$25 mm$^{-1}$.  The FOV is (0$\fdg$5)$^2$.  

A programable multi-object capability is provided by a microshutter array (MSA) placed at the prime focus of the telescope.  The MSA has 43 rows $\times$ 86 columns in the FOV.  An experimental target acquisition system, is designed to operate autonomously with the intent of limiting spectral confusion by closing all but one column on each of the 43 rows.  Each individual shutter subtends a solid angle of $\Omega$ = 12$\farcs4$ $\times$ 36$\farcs$9.

%\begin{center}
\begin{figure}
\includegraphics[bb= 0in 2in 7.5in 11.in,angle=90,width=.5\textwidth]{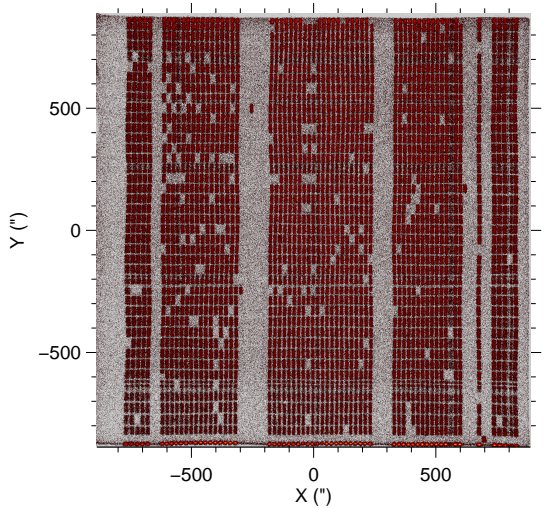}
\caption{Pre-flight image of ``fully opened'' microshutters in \ion{Hg}{1} \lam\ 1849 light.  Approximately 70\% of the shutters are active.
\label{fig1}
}
\end{figure}
%\end{center}

Our flight MSA was derived from prototype versions of the large area arrays developed at Goddard Space Flight Center (GSFC) for use in the Near Infrared Spectrograph (NIRSpec) on the {\it James Webb Space Telescope (JWST)} \citep{Li:2005}.  The shutters are opened with the help of a magnet that passes in front of the array, timed to coincide with a serial stream of opening voltages.  Shutters are closed by reversing the direction of the magnet and applying closing voltages.

The three channel microchannel plate (MCP) detector, custom built by Sensor Sciences,  employs  separate sets of crossed delay-line readout-anodes fed by z-stack MCPs with CsI photocathode inputs.  The deadtime of the pulse counting electronics is 400 ns, however, the maximum countrate of each channel is limited by the telemetry clock rate of the first-in-first-out (FIFO) buffer used to collect the pulse heights and locations in x and y.  The pulse clock rate is 62.5 KHz for the zero-order imaging channel and 125 KHz  each spectral channel.  We  refer to these channels as P1zero, P2minus and P3plus.

The zero-order imager has a short wavelength cutoff defined by the transmission of a CaF$_2$, MgF$_2$ cylindrical doublet lens.  The doublet is included to  correct astigmatism in the imager and limit background counts from the geocoronal \lya.  The slits in the microshutter array have a pitch of 1 mm $\times$ 0.5 mm in the secondary focal plane. A few percent of the shutters are not active due to shorts between columns and/or rows in the array.  These shorts are masked out to prevent drawing excessive current that could potentially damage the array.  Figure~\ref{fig1} is an image of the active microshutters acquired during pre-flight payload qualification testing.   The illumination source was a slow paraxial beam of \ion{Hg}{1} \lam 1849 provided by a penray lamp.   Approximately 70\% of the shutters were active.

%\begin{center}
\begin{figure}
\includegraphics[bb= 1in 1.75in 10.5in 8.5in, width=.5\textwidth]{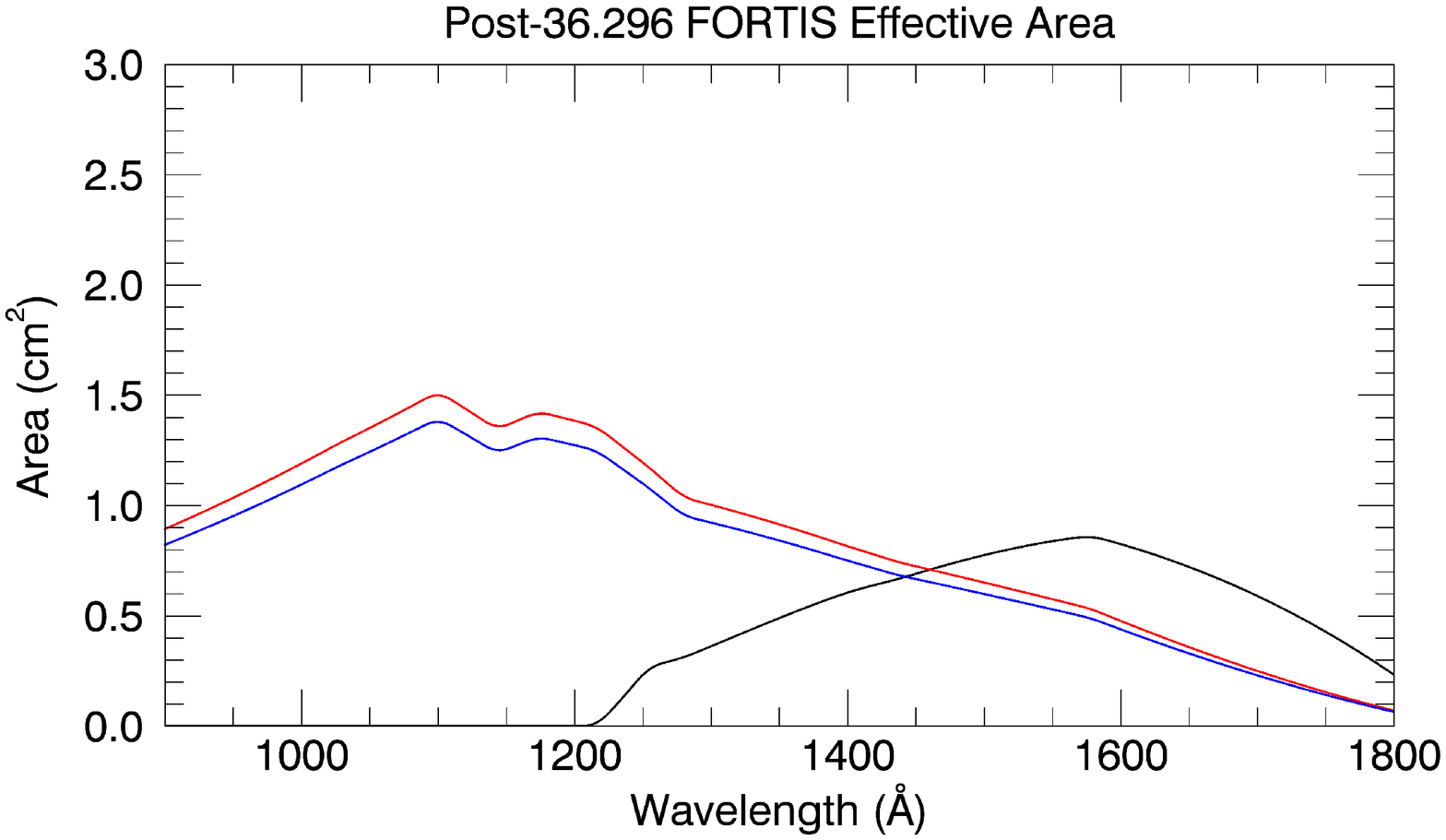}
\caption{Post-flight effective areas for the spectral (red and blue) and imaging (black) channels.  
\label{fig2}
}
\end{figure}
%\end{center}

The effective area of FORTIS was determined from component level efficiency measurements pre- and post-flight made with the Calibration and Test Equipment (CTE) at JHU \citep{Fastie:1975}, following the procedures described by \citet{Fleming:2013}.  The post-flight effective areas of the spectral and imaging channels, shown in Figure~\ref{fig2}, will be used for this work.

\section{Observations}
JHU/NASA sounding rocket 36.296UG was launched from LC-36 at White Sands Missile Range, New Mexico at 04:40 MST on 20 November 2013.  The Black-Brant IX delivery system carried our experimental spectro/telescope, FORTIS, to an apogee of 270 km, providing 395 seconds of exoatmospheric time above 100 km. 

%\begin{center}
\begin{figure*}[ht]
\centering
\includegraphics[bb=0in .5in 6.5in 10.5in, angle=90,width=\textwidth,clip]{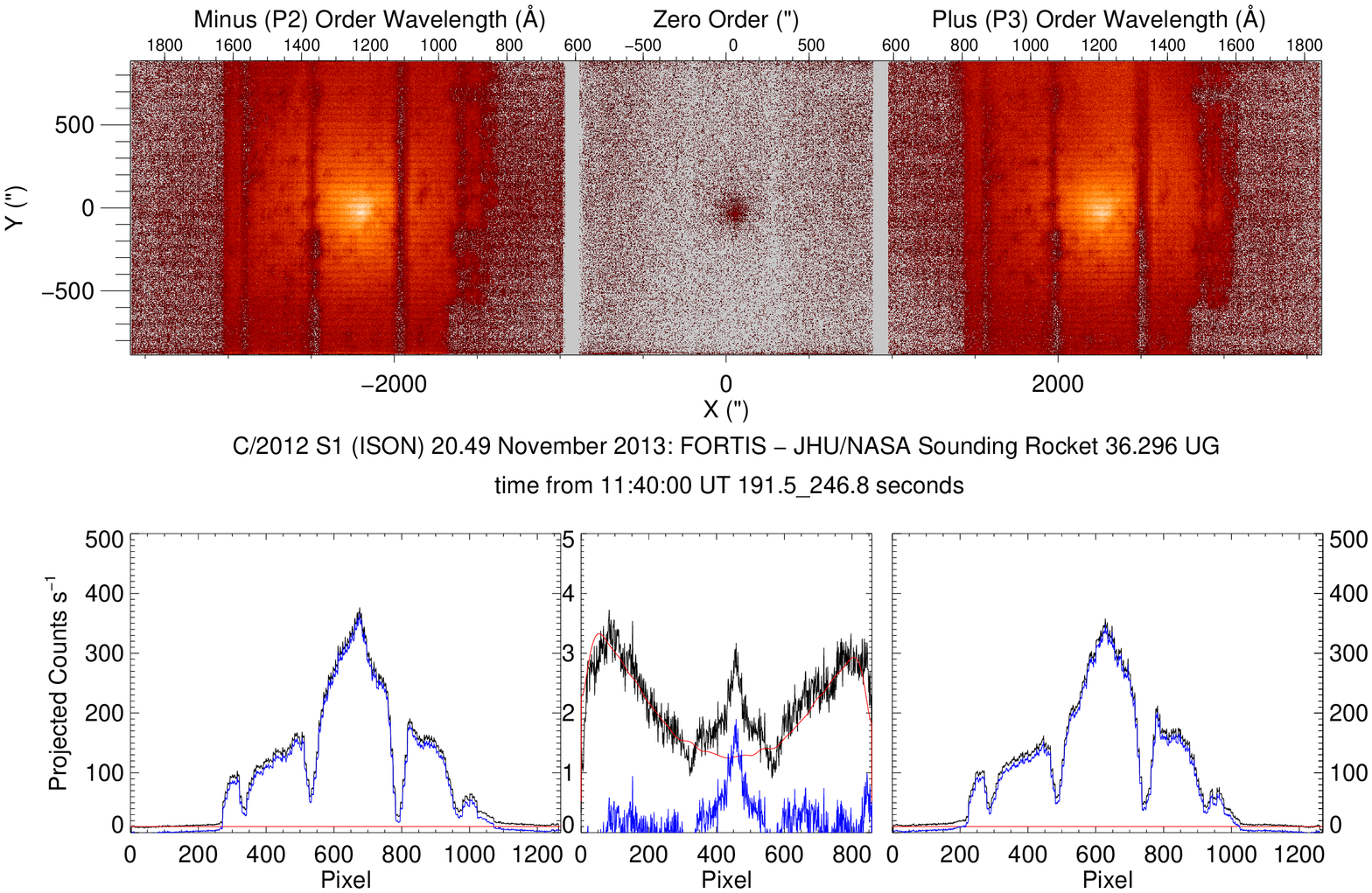}

\includegraphics[bb=0in .5in 6.5in 10.5in, angle=90,width=\textwidth]{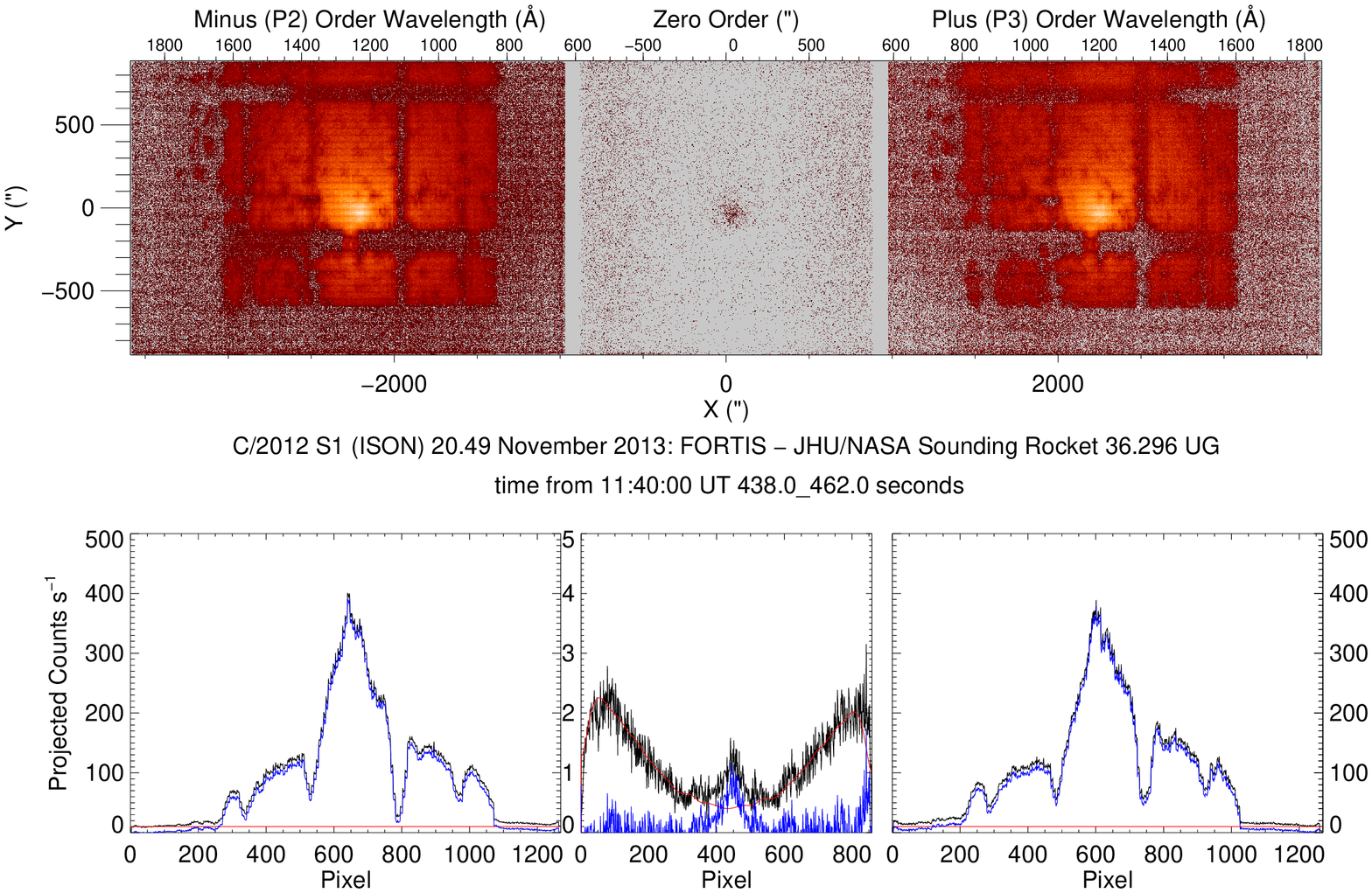}
%\vspace*{.75in}
\caption{Sets of spectral and imaging data.   Pre-apogee sets acquired between T+192 and T+247 seconds on top.  Downleg sets acquired just prior to reentry between T+438 and T+462 seconds on bottom.   X-axes projections are displayed underneath each image to assess background (see text for details).  The zero order images have been multiplied  by a factor of 15 with respect to the spectral channels.   A 1/4 root power law stretch has been applied to all images to elevate faint features.
\label{fig3}
}
\end{figure*}
%\end{center}

In flight, the plan for target acquisition was to observe the comet for 30 seconds in the imaging channel through a fully open MSA, and then deploy a preprogramed slit, in the shape of a K, on the center of brightness.  The location of the center of brightness was to be determined by an on-the-fly peak locating subsystem, serving as an interface between the zero-order imager and MSA.  Unfortunately the preprogramed slit never successfully deployed due to magnet and address timing issues, frustrating our goal of acquiring confusion limited spectral information from selected regions in the coma along and across the sun-comet line in an effort to detect faint volatile species.

Nevertheless zero order images and dispersed spectral images were acquired through a ``mostly open'' MSA, during an $\approx$ 50 second interval ending $\approx$ 40 seconds before apogee.  Three additional attempts to deploy preprogramed slits at roughly 60 second intervals resulted in a ``partially open'' MSA, having similar shutter patterns wherein $\approx$ half of the bottom half of the shutters remained closed.  A large block of open shutters surrounding the comet allowed the extraction of radially averaged profiles from the weaker emissions in the zero-order channel and "slit averaged" profiles from the very strong emissions in the spectral channels.

An additional complication arose in determining the true count in the spectral channels.  During much of the flight the observed rate was saturated at the telemetry sample rate of 125 KHz.  Fortunately, during the second attempt to  deploy a slit the count rate in the ``P2minus'' channel, which is the slightly-less-sensitive of the two, fell below the sample rate to 115 KHz in response to the smaller number of open shutters.  A modest deadtime correction factor of 1.16 was found postflight for the ratio of the true rate to the sampled rate.  The correction was determined in postflight calibrations by taking advantage of the linearity provided by a cesium iodide coated photomultiplier tube (PMT) in response to a steadily increasing illumination from a deuterium lamp.  A  plot of the countrate from the PMT against the countrate of a similarly illuminated FIFO buffered MCP, clocked by telemetry, provided the correction factor.

Representative data acquired pre-apogee and near-reentry are shown in Figures~\ref{fig3}.  The top panels show the on-axis zero-order imaging and off-axis spectral channels. The zero-order imaging has been multiplied by 15 and a 1/4 root stretch has been applied to elevate faint features.  The bottom panels show the projection of the images onto the X-axis, allowing assessment of the background levels to be subtracted.  The zero-order imager has a significant background with respect to the cometary signal.  Initial background measurements were acquired \added{separately} in the zero-order channel before arriving at the comet.  They were used as the basis of a model to account for the background variations in the cometary images during the course of the flight.  The projected background model is shown in red and the result from the subtraction is shown in blue.   

The spectral images are dominated by the intense emission of cometary \lya, with has a peak brightness of 625 Krayleigh. The background levels are quite small in comparison and a small constant level has been subtracted.  We will present evidence in the following section that the \added{cometary} emissions in the imaging channel are dominated by \ion{C}{1} \lam 1657.

\section{Analysis and Results }\label{analnres}

The spatial distribution of cometary emissions provides a means to estimate the gas production rates for its atomic and molecular constituents.  Here we  provide some basic formulae commonly used in the analysis and then move to estimates for the production of water, carbon and carbon monoxide.  We use simple steady state \citet{Haser:1957} models and the \citet{Festou:1981} vectorial models to constrain the production rates.  These models will deviate from the data when variations occur on timescales shorter than the radial scale divided by outflow velocity of a given species. The length scales probed here range from $\sim$ 10$^{3}$ to 10$^{6}$ km.

\subsection{Fluorescence Efficiences (g-factors) \label{gft}}
UV line emission from cometary volatiles arise primarily from absorption of solar flux by resonant (ground state) transitions, producing excited atoms or molecules that then re-emit into 4$\pi$ sr.  Line intensity varies with the square of the distance between the comet and the sun ($r_{h}$).  If there is a coincidence between the resonance absorption and a set of strong line emission within the solar spectrum, as is the case for \ion{C}{1}, then the shape of the solar line profile and the relative velocity between the comet and the sun become another important factor in modulating the fluorescence line intensity \citep{Swings:1941}. \added{We have neglected the Greenstein effect \citep{Greenstein:1958}, wherein the differential velocity of the cometary outflow in the sunward and anti-sunward directions can produce an additional, second-order modulation in fluorescence line intensity. Our solar spectral energy distribution has too low a resolution to sufficiently account for the Greenstein effect.}   

\begin{figure*}[ht]
\centering
\includegraphics[scale=.9]{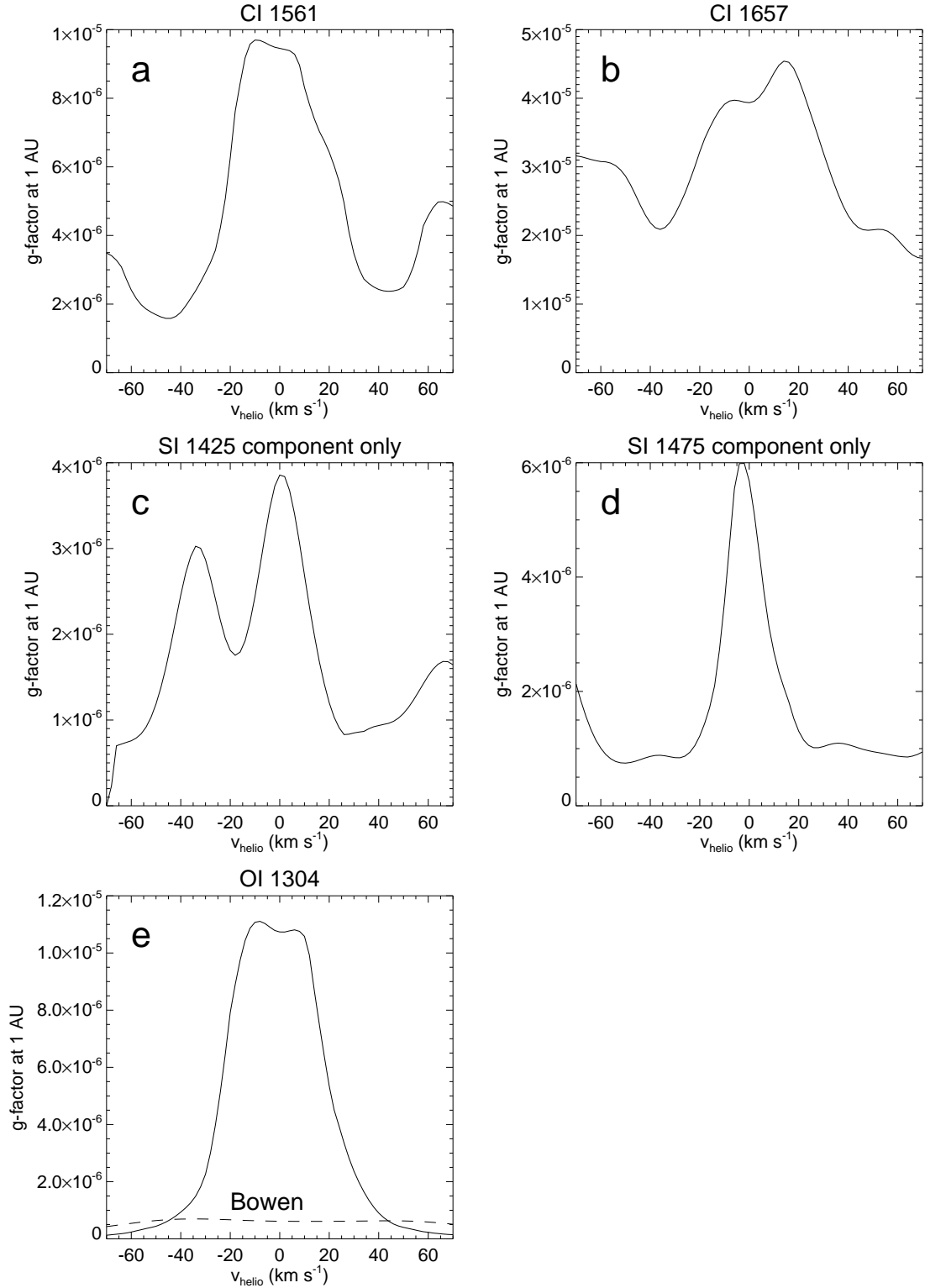}
\caption{Optically thin gfactors accounting for the Swings effect (modulation of the fluorescence line strength by doppler shifting of the solar pumping line in and out of resonance with the absorbing transition in the comet), in the \ion{C}{1} \lam \lam 1561, 1657 (panels a,b),  \ion{S}{1} \lam \lam 1425, 1475 (panels c, d), and \ion{O}{1} \lam 1304 multiplets (panel e). \ion{O}{1} \lam 1304 is pumped directly by the solar multiplet and indirectly through the  coincidence of solar \lyb\ with \ion{O}{1} \lam 1026, which decays though \ion{O}{1} \lam 11287 and \ion{O}{1} \lam 8446 before reaching \ion{O}{1} \lam 1304; a mechanism known as \citet{Bowen:1947} fluorescence. 
\label{fig4}
}
\end{figure*}

The scattering efficiency for a transition, conventionally calculated at 1 AU and commonly known as the ``g-factor'', is given by;

\begin{equation}
g_i (\dot{r_h}) = \lambda_{i}^{2} f_i F_{\odot}(\dot{r_h}) \frac{\pi e^2}{m_e c^2} \frac{A_i}{\Sigma_j A_j}\; {\rm photons\; s^{-1},} 
\end{equation}
where $\lambda_i$\ is the wavelength, $f_i$ is the oscillator strength, $\frac{A_i}{\Sigma_j A_j}$ is the branching ratio for the excited transition (ratio of the $ith$ transition de-excitation rate to the sum of all transitions out of the excited state).  $F_{\odot}(\dot{r_h})$ is the solar photon flux at 1 AU (\photofl) doppler shifted \added{according to} the appropriate heliocentric velocity, $\dot{r_h}$.  The above formula is valid for the optically thin case.  A more sophisticated treatment models the absorbing transition as a Voigt profile integrated over the doppler shifted solar spectrum, so in general the g-factor is also a function of column density.  Optically thin g-factors are shown in Figure~\ref{fig4} for \ion{C}{1} \lam 1561,  \ion{C}{1} \lam 1657,  \ion{S}{1} \lam 1425,  \ion{S}{1} \lam 1475 and \ion{O}{1} \lam 1304, \added{panels a to e respectively}.  We see that \ion{C}{1} \lam 1657 is an order of magnitude stronger than all the others at the most negative velocities. 

In Table~\ref{gfact} we list the optically thin 1 AU g-factors used in this study for \ion{C}{1} \lam 1657, \ion{O}{1} \lam 1304 and \ion{H}{1} \lam 1216 calculated using a heliocentric velocity of $\dot{r_h}~=~-62.7$~km~s$^{-1}$.  We also show representative g-factors for the strongest two CO A-X bands (1-0) at 1510 \AA\ and (2-0) at 1478 \AA.  The CO bands are pumped by continuum photons.  A solar spectral energy distribution \added{associated} with a moderately active F10.7 \added{cm} flux of $\sim$ 150 sfu (solar flux unit),\footnote{1 sfu = 10$^4$ Jy = 10$^{-19}$ erg cm$^{-2}$ s$^{-1}$ Hz$^{-1}$} as appropriate to the time of observation,  was used for the \ion{C}{1} \lam 1657 and \ion{O}{1} \lam 1304 g-factor calculations.  The \ion{H}{1} \lam 1216 g-factor was interpolated from Table 1 of \citet{Combi:2014}.

%\begin{center}
%\floattable
\begin{deluxetable}{lcc}
\tablecaption{\bf g-factors at 1 AU and $\dot{r}_h = -62.7$ km s$^{-1}$ \label{gfact} }
\tablecolumns{3}
\tablewidth{0pt}
\tablehead{\colhead{Species}	& \colhead{Wavelength} &	\colhead{g-factor} }
\startdata
\ion{C}{1} 						&1657 	&3.2 $\times$ 10$^{-5}$ \\
CO \coApi\ -- \coXsig  (1-0)\tablenotemark{a} 	&1510	& 1.9  $\times$ 10$^{-7}$	 \\
CO \coApi\ -- \coXsig  (2-0) \tablenotemark{a}	&1478	& 1.8  $\times$ 10$^{-7}$	  \\
CO \coApi\ -- \coXsig  (all bands)\tablenotemark{a} & 1280 -- 1800 & 1.5  $\times$ 10$^{-6}$ \\
\ion{S}{1} 						&1474	& 	1.1 $\times$ 10$^{-6}$ \\		
\ion{O}{1}						&1302	& 	6.0 $\times$ 10$^{-7}$	\\
\ion{H}{1}\tablenotemark{b}						&1216	&   2.2 $\times$ 10$^{-3}$ \\
\enddata
\tablenotetext{*}{CO bands are pumped by solar continuum}
\tablenotetext{b}{Taken from \citet{Combi:2014}} 
\end{deluxetable}
%\end{center}
%\vspace*{-.5in}

The scattering takes place into 4$\pi$ sr, so the brightness (in rayleighs)\footnote{Rayleigh $\equiv$ 10$^6$/(4$\pi$) photons cm$^{-2}$ s$^{-1}$ sr$^{-1}$.} of a cometary emission line at arbitrary heliocentric distance is given by;

\begin{equation}
B_i = \frac{\bar{N}\; g_i}{10^{6}\; r_{h}^2},
\end{equation}
\noindent where $\bar{N}$ is the mean column density (cm$^{-2}$).   If all the emission from the comet can be contained within an aperture whose solid angle $\Omega$ subtends an area $A_{aper}$ = $\Omega \Delta^2$ at the comet then the production rate is simply;

\begin{equation}
Q =\frac{ \bar{N}\; A_{aper}}{\tau},
\end{equation}
\noindent where $\tau_i$ is the lifetime of the species in question.   In general, the photodissociation lifetime of a species is proportional to the incident solar flux, so $\tau_i$ will scale with heliocentric distance $r_{h}^{2}$.  

In cases where the emissions are extended with respect to the aperture it is common to model parent species, like CO, as a steady-state outflow from the nuclear regions of the coma, having a constant velocity $v$ and an exponential scale length \citep{Haser:1957}. The number density as a function of radius $r_c$ is given as;

\begin{equation}
n = \frac{Q}{4 \pi v r_c^2} \exp{(-\beta\;r_c)},
\end{equation}
where $\beta = (v\tau)^{-1}$ is the inverse scale length.  This model can be projected onto the line of sight to yield a column density, so a plot of the brightness as a function of radius yields a column density profile from which the product rates for the volatile species can be determined. A more sophisticated vectorial model \citep{Festou:1981} accounts for the production of daughter products, emitted isotropically in the rest frame of the dissociating parent species.

\subsection{Water Production Rate from \lya\ Image}

Water (H$_2$O) is well known to be the dominant volatile constituent of comets.  Upon sublimation the parent molecule H$_2$O dissociates into its atomic and molecular constituents, referred to as daughter products, under the influence of solar photons and solar wind particles. \citet{Budzien:1994} have provided a thorough discussion of the various water destruction channels, including techniques to account for varying levels of  extreme- and far-UV variation throughout the solar cycle.  \citet{Combi:2005} have pointed out that, in addition to providing an estimate for the water production rate, the spatial distribution of \lya\ also provides information on the velocity distribution of the H daughter.  In the  analysis provided here we will neglect contributions to the H production from sources other than water and its direct dissociation products.  

\begin{figure}[ht]
\includegraphics[angle=90,width=.5\textwidth]{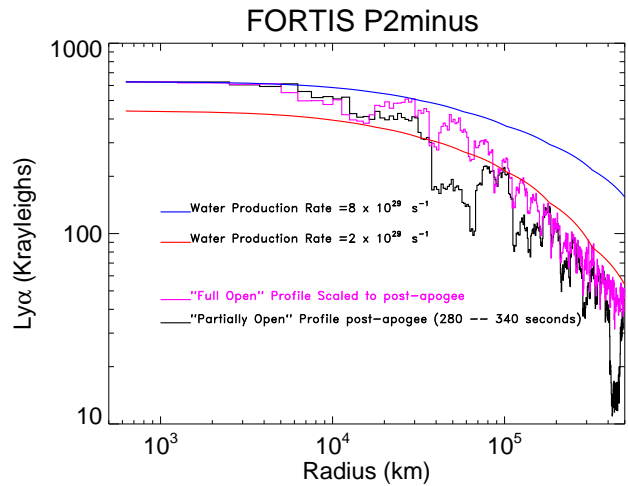}
\caption{\lya\ profiles summed over a 41$\farcs$25 wide slit centered on the comet. The profile in red was acquired pre-apogee between T+190 and T+240 seconds when the MSA was in an fully opened state.  In this state the count rate was saturated at the maximum sample rate (125 kHz).   The profile in black was acquired post-apogee between T+280 and T+340 seconds when an attempted addressing of a slit resulted in a partially opened MSA.  This lowered the detected count rate below that of the sample rate, allowing an accurate deadtime correction to be made. The peak of the post-apogee profile is our most accurate estimate of the central Lya brightness as sampled over 2$\arcsec$.  The pre-apogee profile has been scaled so its peak matches the post-apogee observation.    The shape of the pre-apogee profile has fewer closed shutters, providing a smoother sample of the profile in comparison to the post-apogee profile. The true profile is best represented by the upper envelope of each profile.  Comparison to steady state Haser models implies a water production rate between $Q_{H_2O}$ 8 $\times$ 10$^{29}$ s$^{-1}$ at small radii, while the larger radii are better matched by a lower rate of 2 $\times$ 10$^{29}$ s$^{-1}$.  The disagreement with the steady state models suggests a strongly increasing in the water production rate.  
\label{fig5}
}
\end{figure}

\begin{figure*}
\centering
\includegraphics[bb= 0in 0in 3in 11in,angle=90,width=\textwidth]
{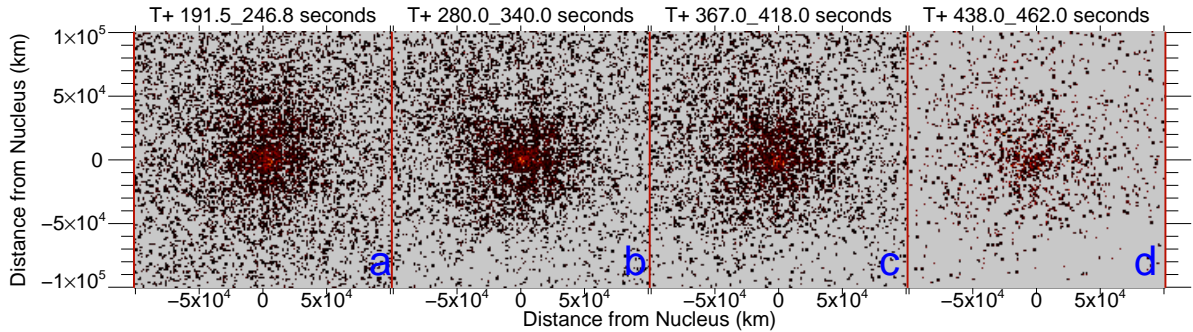}
\caption{Zero order image zoom-in to $\pm$ 1e5 km (332$\arcsec \times$ 332$\arcsec$). The observation intervals, post launch, are indicated above each image. 
\label{fig6}
}
\end{figure*}
In Figure~\ref{fig5} we show two profiles extracted from a 20 pixel (41$\farcs$25 - two shutters wide) region centered on the brightest region of the P2minus detector and extending in the anti-sun direction.  The profile shown in black was acquired post-apogee when the MSA was in a ``partially opened'' state and the count rate was not saturated.  The jagged shape of the profile is due to closed shutters along the extraction direction.   The profile shown in red was acquired pre-apogee when the MSA was in a ``mostly opened'' state but the count rate was saturated.  The overall shape is less affected by closed shutters.  The red profile has been shifted to match the core region of the unsaturated profile where the shutters are fully opened.

We have over plotted \lya\ radial profiles for water production rates of 8  $\times$ 10$^{29}$ and 2 $\times$ 10$^{29}$ s$^{-1}$,  derived from steady state Haser models modified to include compensation for saturated radial profiles that become optically thick towards the center of the coma.  The higher rate is a reasonable match to the upper envelope of the core region at radii $\lesssim$ 5 $\times$ 10$^4$ km, while the lower rate matches the upper envelope towards the outer regions at radi $\gtrsim$  10$^5$ km.  This is suggestive of an increasing water production rate in apparent agreement with that observed by \citet{Combi:2014}, who found $Q_{H_2O} =$  3.8 $\times$ 10$^{29}$ s$^{-1}$ on 19.6 November, and 19.4 $\times$ 10$^{29}$ on 21.6 November\added{, albeit from a large aperture observation}.  \added{Our water production observation is most closely bracketed by those from \citet{DelloRusso:2016}, which were derived from the small aperture of IRTF/CSHELL on the nights of 19 and 20 November 2013.  Starting at 19.71 November they found a water production rate of $Q_{H_2O} =  2.4 \pm 0.1 \times$ 10$^{29}$ s$^{-1}$ , which steadily increased to   $Q_{H_2O} =  4.4 \pm 0.3 \times$ 10$^{29}$ s$^{-1}$ by 19.96 November.  Between 20.70 and 20.88 November they found an average value of $Q_{H_2O} =  3.7 \pm 0.4 \times$ 10$^{29}$ s$^{-1}$. }  

All these  observations are well bracketed by the water production rates found by  \citet{DiSanti:2016}, using IRTF/CSHELL.   They quote $Q_{H_2O} =  1.6 \pm 0.1 \times$ 10$^{29}$ s$^{-1}$,   4.3 $\pm 0.3 \times$ 10$^{29}$ s$^{-1}$,  9.9 $\pm 0.5 \times$ 10$^{29}$ s$^{-1}$ and 4.1 $\pm 0.2 \times$ 10$^{29}$ s$^{-1}$, on November 18.7, 19.9, 22.7 and 23.0 respectively.

\subsection{Carbon Production}
\label{carbon}
In Figure~\ref{fig6}  we show background subtracted count rate images with linear scaling from the zero-order channel covering a 332$\arcsec\ \times$ 332$\arcsec$ region ($\pm$ 10$^5$ km)$^2$.  The zero-order imaging bandpass, ranging over $\sim$ 1300 to 1800 \AA, is sensitive to the emission from a number of cometary species.  The 1 AU fluorescent efficiencies (g-factors) listed in Table ~\ref{gfact} show that \ion{C}{1} \lam 1657  has the highest g-factor followed by the band sum of CO, \ion{S}{1} \lam 1475 and \ion{O}{1} \lam 1302 respectively.  Here we present evidence in support of \ion{C}{1} \lam 1657 as the dominant source of emission in the zero-order images.

Observations using the Cosmic Origins Spectrograph (COS) on the {\it Hubble Space Telescope} on 01 November 2013, when the heliocentric velocity was $-42$ km s$^{-1}$, found a \ion{S}{1} \lam 1425 line that was $\approx$ 5 times stronger than the \ion{S}{1} \lam 1475 and comparable in strength to \ion{C}{1} \lam 1657 \citep{Weaver:2014}.  However, as shown in Figure~\ref{fig4}\added{c}, the g-factor for \ion{S}{1} \lam 1425 has a strong dependence on heliocentric velocity, dropping by a factor of 3 at $-62.7$ km s$^{-1}$, \added{with respect to that at $-42$ km s$^{-1}$,}  and is below that of \ion{S}{1} \lam 1475 \added{(panel d)}.  We further note that the COS aperture is only 2$\farcs$5 in diameter, comparable to our pixel and much smaller than the extractions shown in Figure~\ref{fig6}.  Cometary sulfur emissions typically extend over a much narrower angular extent in comparison to carbon \citep[e.g.][]{McPhate:1999},  so the \ion{C}{1} \lam 1657 intensity measured by COS samples only a small fraction of the flux available.  We conclude that sulfur contributions in our zero-order image are likely to be $\sim$ 10\% that of carbon.   The band integrated g-factor for CO, is 1.4 times that of \ion{S}{1} \lam 1475 \added{(Table~\ref{gfact}),} and like sulfur has a narrow angular distribution in comparison to carbon, hence we expect it to be similarly weak. 

Oxygen is a strong byproduct of water dissociation.  However, its g-factor at $-62.7$ km s$^{-1}$ is \added{a factor of} $\sim$ 50 smaller than the carbon line.  Moreover, the geocentric velocity of the comet is only $-4.5$ km s$^{-1}$ which leads to strong attenuation of \deleted{this line} \added{\ion{O}{1} \lam 1302} by atomic O in the thermosphere where slant column densities at the observation angle of 89$\degr$ from zenith\added{,} range from 10$^{19}$ to 10$^{16}$ cm$^{-2}$ at altitudes between 100 and 300 km.  

\begin{figure*}
\centering
\includegraphics[bb= 0in 0in 7.in 11.in,angle=90,width=\textwidth,clip]{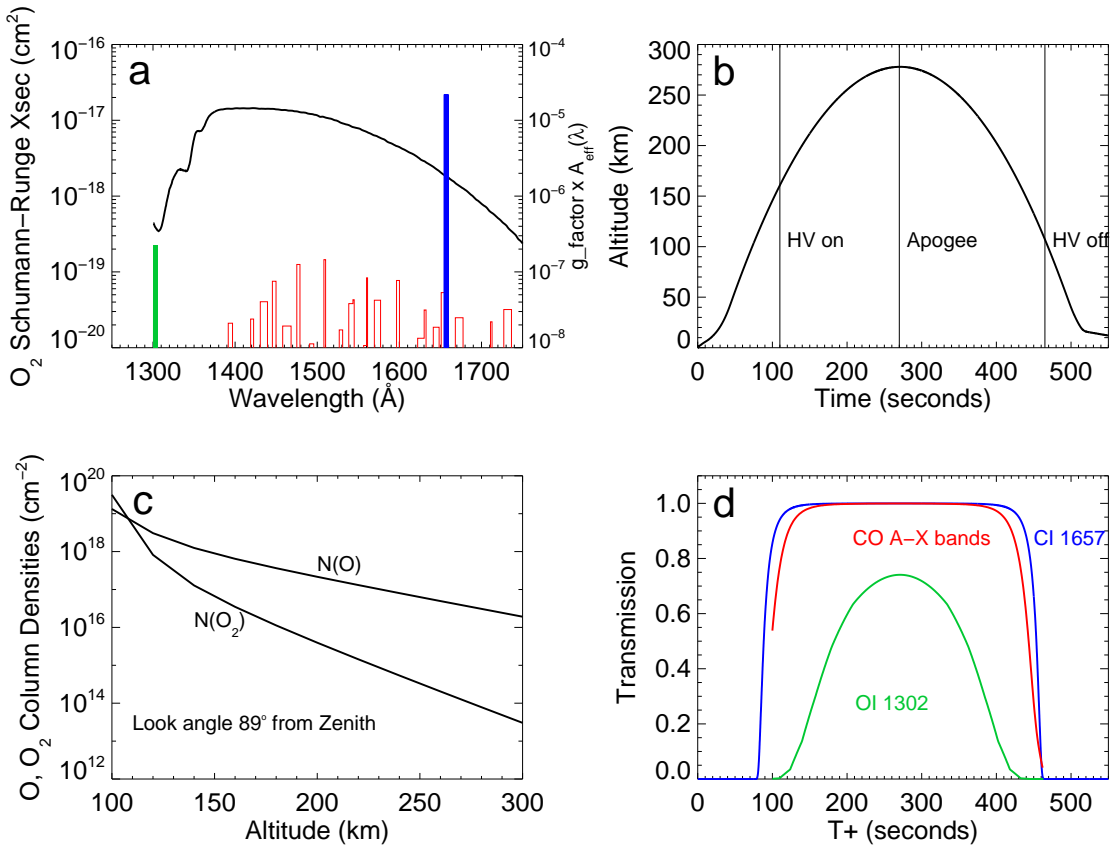}
\caption{\deleted{Top left} \added{a} -- Molecular oxygen Schumann-Runge continuum absorption cross section in black. The g-factors at 1 AU, multiplied by the zero-order effective area in Figure~\ref{fig2}, are shown in red for the CO A-X band transitions, in blue for \ion{C}{1} \lam\ 1657  and in green for \ion{O}{1} \lam\ 1302.  \deleted{Top right} \added{b} -- Rocket altitude as a function of time.  \deleted{Bottom left} \added{c} -- Slant angle column densities for O and O$_2$ at a zenith angle of 89$\degr$.  \deleted{Bottom right} \added{d} -- Atmospheric transmission as a function of time for \ion{C}{1} \lam 1657 (blue), the g-factor weighted CO emission, \ion{O}{1} \lam 1302.    
\label{fig7}
}
\end{figure*}

We can further discriminate between carbon and other potential emitters in the zero-order images by monitoring the count rate as the telescope descends into Earth's atmosphere. Molecular oxygen (O$_2$) absorption has a strong dependence on wavelength, which will selectively attenuate cometary emissions from different atomic and molecular species at different rates on the downleg portion of the flight.  We have modeled the expected attenuation as a function of time during the flight for \ion{C}{1} \lam\ 1657, \ion{O}{1} \lam\ 1302 and the CO A-X band.  A summary of the components of this model is shown in Figure~\ref{fig7}\added{abcd}.

In \deleted{the top left panel}\added{Figure~\ref{fig7}a} we show the O$_2$ Schumann-Runge continuum absorption cross section in black. The g-factors at 1 AU, multiplied by the zero-order effective area in Figure~\ref{fig2}, are shown in red for the CO A-X band transitions, in blue for \ion{C}{1} \lam\ 1657, and in green for \ion{O}{1} \lam\ 1302.  The altitude of the telescope as a function of time is shown in \deleted{the top right panel}\added{Figure~\ref{fig7}b}.  In \deleted{the bottom left panel}\added{Figure~\ref{fig7}c} we show the slant column densities of O$_2$ and O as a function of altitude.  In \deleted{the bottom left panel}\added{Figure~\ref{fig7}d} we show the transmission profiles from each emission source as a function of time.

\begin{figure}[ht]
\includegraphics[angle=90,width=.5\textwidth]{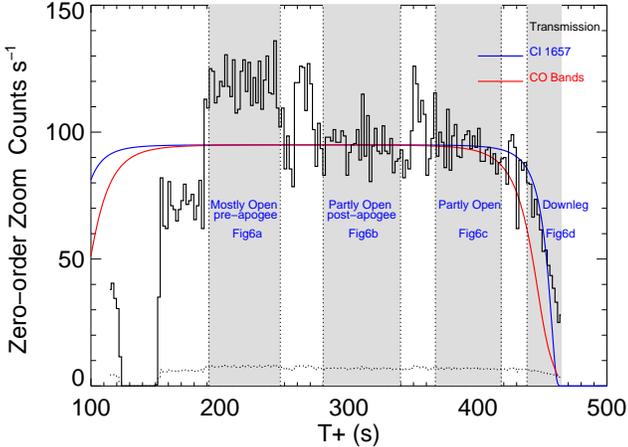}
\caption{Zero-order count rate over $\pm$ 10$^5$ km region centered on the nucleus in  black \added{histogram}.  The vertical dotted lines mark periods where the total number of open shutters changed in response to target acquisition and attempts to deploy pre-programed slits.  \added{The grey shaded regions mark times over which the images in Figure~\ref{fig6}abcd were extracted.} The  atmospheric transmission  at the wavelength of the \ion{C}{1} \lam 1657 emission is shown in blue.  The atmospheric transmission for the CO bands is shown in red.  The count rate at reentry is consistent with \ion{C}{1} \lam 1657 as the dominant emitting species in the zero-order channel.  \added{The dotted line at the bottom indicates the 1-$\sigma$ statistical error for the count rate.}
\label{fig8}
}
\end{figure}

\begin{figure}[hb]
\includegraphics[angle=90,width=.5\textwidth]{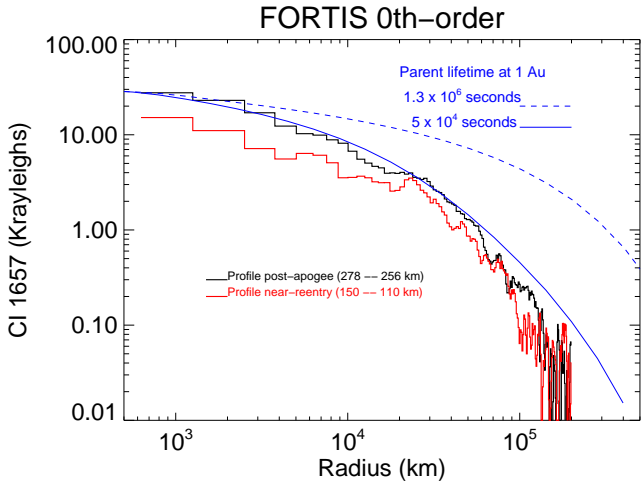}
\caption{Zero order profile averaged over concentric circles surrounding the comet.  The flux in this profile is dominated by emission from of \ion{C}{1} \lam 1657 as discussed in \S~\ref{carbon}.  Black is the profile acquired post apogee.  Red is the profile near reentry when the attenuation due to the atmosphere is strongest.  Overplotted in blue is a vectorial model for the steady state production of carbon from a parent molecule at the rate of \deleted{3}\added{4} $\times$ 10$^{28}$ s$^{-1}$ and a 1 AU lifetime of 5 $\times$ 10$^4$ seconds.  The dashed blue profile was computed assuming a lifetime typical of that for C  produced by a CO parent 1.3 10$^6$ seconds.
\label{fig9}
}
\end{figure}

\begin{figure*}
\centering
\includegraphics[bb= 3.5in 1in 7.in 10.in,angle=90,width=.95\textwidth]{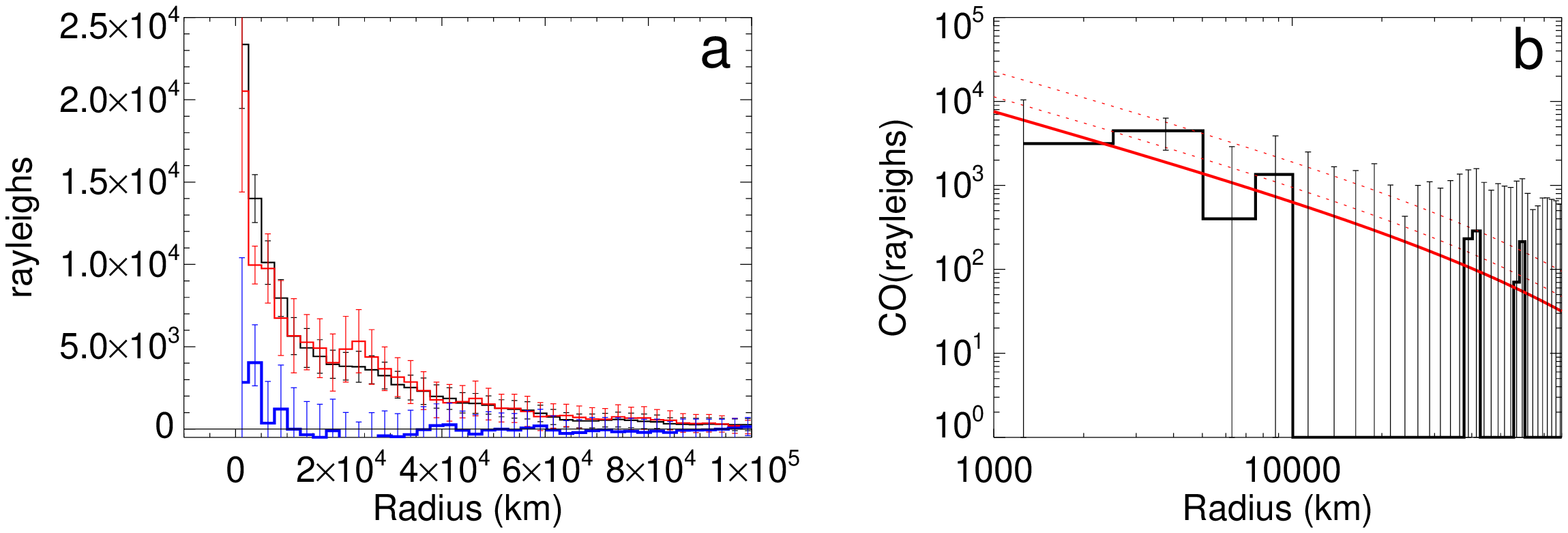}
\caption{Profile subtraction of post-apogee image from downleg image is shown in the left panel on a linear scale.  The 0 level is indicated with a horizontal line. The result is in blue. \added{One standard deviation error bars are over plotted.} In the right panel we over plot the resulting CO profile with a Haser model for CO sublimating from the surface at the rate of 5 $\times$ 10$^{28}$ s$^{-1}$. \deleted{A $\pm$50\% error envelope is indicated.}\added{Error bars, added in quadrature, are over plotted. Upper limit envelops are shown at 1.5 and 3 $\times$ the model profile.}  
\label{fig10}
}
\end{figure*}

In Figure~\ref{fig8} we show the zero-order count rate over $\pm$ 10$^5$ km region centered on the nucleus in black.  The vertical dotted lines mark times when total number of open shutters were changed in response to attempts to deploy a slit.  The transmission curve as a function of time for \ion{C}{1} \lam\ 1657 is an excellent match to the count rate during the period of reentry.  This is strong evidence for carbon as the dominant source of emission in these images.  The lack of any sort of \deleted{parabolic variating}\added{parabolically varying} component in this rate is \deleted{complementary}\added{an} indicator that \ion{O}{1} \lam\ 1302 is not present at a significant level.

In Figure~\ref{fig9} we plot \added{as a black histogram} the radial profile of the zero-order emission \deleted{as} averaged over \deleted{concentric circles around}\added{annuli centered on} the pixel of peak brightness \deleted{for}\added{in} the image acquired post apogee \deleted{as a black histogram.}\added{(Figure~\ref{fig6}b)}.   The radial profile for the image acquired at the end of the downleg \added{(Figure~\ref{fig6}d)} is plotted in red.  The later profile has a less pronounced peak.  In \S~\ref{CO} we will use the difference of these two profiles to constraint the CO production rate.

We overplot vectorial models representative of carbon as produced by a parent with 1 AU lifetimes of 1.3 $\times$ 10$^6$  and 5 $\times$ 10$^4$ seconds as dashed and solid blue lines respectively.  The former lifetime is that expected from a CO parent.  It clearly does not fit the observation.  The later lifetime provides a good fit to the observation, but leads to the conclusion that a parent molecule with a much short lifetime  than CO is responsible for the C  production.  $Q_{C} \approx$ \deleted{3}\added{4}~$\times$~10$^{28}$~s$^{-1}$, assuming a parent lifetime of 5 $\times$ 10$^4$ seconds.

\subsubsection{CO Production Constraint \label{CO}}

Parent molecular species sublimating directly from the \deleted{cometary body}\added{nucleus} with optically thin column densities, like CO, are point sources as viewed by Earth bound telescopes, exhibiting a sharp peak at the center of an image.  Our atmospheric transmission calculations, along with the zero-order count rate observation (Figure~\ref{fig8}),  indicate that the downleg image should be mostly devoid of CO emission and should contain only \ion{C}{1} \lam 1657 emission, albeit somewhat attenuated.  The difference between \deleted{a}\added{the} profile acquired post-apogee \added{(Figure~\ref{fig6}b), which has \ion{C}{1} \lam 1657 and possibly some CO at the center,}  and that acquired on the down leg \added{(Figure~\ref{fig6}d), which has only \ion{C}{1} \lam 1657,} allows us to place a limit on the level of CO production.

In \deleted{the left panel of}Figure~\ref{fig10}\added{a} we plot the zero-order image profile acquired post-apogee in black with the downleg profile in red.  We have \deleted{scaled}\added{multiplied} the downleg profile by a factor of 1.6 to account for atmospheric attenuation of \ion{C}{1} \lam 1657, which provides a good match to the wing of the post-apogee profile at large radius.   The difference between the two profiles is shown in blue.  In \deleted{the right panel}Figure~\ref{fig10}\added{b} we show, on a log-log plot, the differenced radial profile in black along with a simple Haser model \added{in red} for CO sublimating from the comet \deleted{at}\added{with} a production rate of \deleted{$Q_{CO} < $ 5 $\times$ 10$^{28}$ s$^{-1}$}\added{$Q_{CO} = 5^{+1.5}_{-7.5} \times 10^{28}$ s$^{-1}$ with 1$\sigma$ errors derived from photon statistics}.   This can only be considered an upper limit as the enhancement is slight and dominated by a single resolution element.

\section{Conclusions}

We find a \lya\ radial profile that is not well matched by Haser models with steady state water production.  At small radii the emission is consistent with a production rate of $Q_{H_2O} \sim$ 8 $\times$ 10$^{29}$ s$^{-1}$, while at large radii the profile is better matched by a lower production rate of $\sim$ 2 $\times$ 10$^{29}$ s$^{-1}$.  This suggests that at the time of our observation, 20.49 November 2013, ISON was undergoing a strong increase in water production on linear scales of 10$^3$ to 5 $\times$ 10$^{5}$ km.   The overall increase is in general agreement with the SWAN observations by \citet{Combi:2014}, and well bracketed by the \citet{DiSanti:2016} \added{and \citet{Dello Russo:2016}} water production determinations. However, it is somewhat at odds with the SWAN daily average, which shows a slight downward trend in the water production rate at the time of our observation. The SWAN daily average was calculated from a time resolved model that accounts for the photodissociation kinetics and thermalization processes from various H parent species that effect the \lya\ brightness profile on scales $\ge$ 1\degr.  This dimension is significantly larger than  our entire FOV of (30$\arcmin$, $\sim$ 10$^{6}$ km), offering a potential explanation for the discrepancy, and suggesting our close in look may offer insight into the dissociation processes during the disruption event.

The flux in the imaging channel is consistent with \ion{C}{1} $\lambda$ 1657 emission.  We find a carbon production rate of $\sim$ \deleted{3}\added{4} $\times$ 10$^{28}$ s$^{-1}$ with a parent lifetime of $\sim$ 5 $\times$ 10$^{4}$ seconds.  This lifetime is shorter than expected from CO, implying CO is not \added{a} dominant source of carbon in coma. \citet{Lim:2014} and \citet{Morgenthaler:2011} came to similar conclusions regarding comets C/2001 Q4 (NEAT) and C/2004 Q2 (MACHHOLZ), although our lifetime is considerably shorter \added{than} that inferred for those two comets.  Our production rate of carbon with respect to water is  C/H$_2$O $\approx$ 5\%.

We have taken advantage of the variable transmission to select far-UV wavelengths offered by molecular oxygen in the Earth's atmosphere to constrain the production rate of CO sublimating from the surface of ISON to be $Q_{CO} \lesssim $ 5 $\times$ 10$^{28}$ s$^{-1}$.  The upper limit on the ratio of carbon monoxide to water is $<$ 6\%.  This upper limit is considerably larger than those derived from COS observations.  They found $Q(CO)$ = 3 $\times$ 10$^{26}$ s$^{-1}$ and 2.7 $\times$ 10$^{26}$ s$^{-1}$ on 22 October and 01 November respectively, when the comet was at heliocentric distances of $\approx$ 1.2 and 1.0 AU \citep{Weaver:2014}.  Deconvolved daily averages of the water production rates derived from the SWAN \lya\ imager \citep{Combi:2014} indicate $Q(H_{2}O)$ = 1.6 $\times$ 10$^{28}$ s$^{-1}$ and  2.2 $\times$ 10$^{28}$ s$^{-1}$ around those dates, suggestive of a gradual decrease in the CO/H$_2$O ratios of 1.9 to 1.2\%, in line with the non-steady evolution that characterized ISON's ingress.

In future work we intend to examine our data in the context of nearly contiguous far-UV spectral observations acquired over 19 to 21 November 2013 made by Mercury Atmospheric and Surface Composition Spectrometer (MASCS) on NASA's {\it MESSENGER} spacecraft to further investigate the water production variability  and to place more stringent limits on the CO production, during this extremely volatile period.

\acknowledgments
The authors would like to acknowledge the sacrifices made by the personnel associated with the NASA Sounding Rocket Program Office, their Contractors, the Navy Launcher Team and the Army Range Control at White Sands Missile Range, all of whom showed exemplary dedication in carrying out this time critical mission.  We would also like to acknowledge the innumerable, essential and critical contributions of our JHU project engineer, Russell Pelton, in providing support to this mission.  Funding for this work was provided to the Johns Hopkins University through NASA sounding rocket grants numbered NNX11AG54G and NNX14AI78G.

\facilities{Wallops Flight Facility, White Sands Missile Range}

\listofchanges

\end{document}